\documentclass[%
prx,
 amsmath,amssymb,
 aps,
twocolumn,
longbibliography,
showkeys
]{revtex4-1}

\usepackage{graphicx}
\usepackage{epstopdf}
\usepackage{dcolumn}
\usepackage{bm}
\usepackage{color}
\usepackage[mathlines]{lineno}
\usepackage[breaklinks=true,colorlinks=true,linkcolor=blue,urlcolor=blue,citecolor=blue]{hyperref}
\usepackage{ragged2e}
\usepackage{enumitem}
\usepackage{bibunits}
\usepackage{amsthm}
\usepackage{lineno}

\newtheorem{theorem}{Theorem}
\newcommand{\lr}[1]{\left(#1\right)}
\newcommand{\lrs}[1]{\left[#1\right]}

\renewcommand\thefigure{\arabic{figure}}

\begin{document}


\title[]{Exit options sustain altruistic punishment and decrease the second-order free-riders, but it is not a panacea}

\author{Chen Shen$^{1,2}$}
\author{Zhao Song$^3$}
\author{Lei Shi$^2$}
\email{shi\_lei65@hotmail.com}
\author{Jun Tanimoto$^1$}
\author{Zhen Wang$^{3,4}$}
\email{w-zhen@nwpu.edu.cn}

\affiliation{
\vspace{2mm}
\mbox{1. Faculty of Engineering Sciences, Kyushu University, Kasuga-koen, Kasuga-shi, Fukuoka 816-8580, Japan}
\mbox{2. School of Statistics and Mathematics, Yunnan University of Finance and Economics, Kunming 650221, China}
\mbox{3. School of Mechanical Engineering,Northwestern Polytechnical University, Xi'an 710072, China}
\mbox{4. School of Artifcial Intelligence, OPtics and ElectroNics (iOPEN),}\\
\mbox{Northwestern Polytechnical
University, Xi'an 710072, China}
}

\date{\today}

\begin{abstract}
  Altruistic punishment, where individuals incur personal costs to punish others who have harmed third parties, presents an evolutionary conundrum as it undermines individual fitness. Resolving this puzzle is crucial for understanding the emergence and maintenance of human cooperation. This study investigates the role of an alternative strategy, the exit option, in explaining altruistic punishment. We analyze a two-stage prisoner's dilemma game in well-mixed and networked populations, considering both finite and infinite scenarios. Our findings reveal that the exit option does not significantly enhance altruistic punishment in well-mixed populations. However, in networked populations, the exit option enables the existence of altruistic punishment and gives rise to complex dynamics, including cyclic dominance and bi-stable states. This research contributes to our understanding of costly punishment and sheds light on the effectiveness of different voluntary participation strategies in addressing the conundrum of punishment.

\end{abstract}


\maketitle

\section*{Introduction}
Costly punishment is observed in many animal species, including humans~\cite{clutton1995punishment,west2007social,henrich2006costly}. Humans, unlike other animals, often exhibit altruistic behavior by punishing individuals who have harmed others, even at their own expense~\cite{fehr2002altruistic,henrich2006costly}. However, the emergence and maintenance of altruistic punishment present an evolutionary puzzle because costly punishment is unlikely to evolve through natural selection. Costly punishment reduces the payoff for both the punisher and the punished. If survival favors the fittest, individuals who cooperate without punishing are better off than punishers, and defectors should eventually dominate the population. Therefore, understanding the evolution of costly punishment is crucial to studying human cooperation.

Numerous experimental and theoretical studies have been conducted to investigate the conundrum of punishment, following the pioneering work of Fehr and G\"{a}chter~\cite{fehr2000cooperation,fehr2002altruistic}. Experimental studies have explored various aspects of punishment, including the cost-effectiveness of costly punishments~\cite{egas2008economics,nikiforakis2008comparative}, the possibility of perverse punishments~\cite{nikiforakis2008punishment}, and the use of monetary punishments~\cite{masclet2003monetary}, among others~\cite{sefton2007effect}. A comprehensive review of these studies can be found in Chaudhuri's work~\cite{chaudhuri2011sustaining}. Theoretical studies have focused on solving the conundrum of punishment through spatial structures~\cite{helbing2010evolutionary,szolnoki2017second,helbing2010punish,perc2012self} or various social mechanisms, including reputation~\cite{brandt2003punishment,ohtsuki2009indirect}, group selection~\cite{boyd2003evolution,saaksvuori2011costly,gavrilets2014solution}, prior commitment~\cite{molleman2019people,han2016emergence}, voluntary participation~\cite{brandt2006punishing,hauert2007via,fowler2005altruistic}, and others~\cite{eldakar2008selfishness}. Theoretical models that consider spatial structure have shown that direct competition between altruistic punishment and individuals who cooperate but never punish can be avoided, addressing the issue of second-order free-riders~\cite{brandt2003punishment}. These results remain valid even when considering detrimental forms of punishment~\cite{szolnoki2017second}.

In the absence of spatial structure, the mechanisms behind the emergence of altruistic punishment in social scenarios differ. Reputation can solve the conundrum of punishment by allowing altruistic punishers to gain indirect benefits through maintaining a good reputation of punishing wrongdoers~\cite{hilbe2012emergence}. Group selection, on the other hand, explains the emergence of altruistic punishment by suggesting that groups with higher levels of cooperation and punishment may have a selective advantage over groups with lower levels of these traits~\cite{boyd2003evolution}. Prior commitment provides a mechanism for the emergence of altruistic punishment by allowing individuals to signal their willingness to bear the costs of punishing and help establish a norm of cooperation that discourages free-riding behavior~\cite{boyd2010coordinated}. In cases where participation in the game is not mandatory but players have the freedom to choose whether or not to participate, voluntary participation can establish altruistic punishment in both finite and infinite populations~\cite{hauert2007via,mathew2009does,fowler2005altruistic,brandt2006punishing,sasaki2012take}. In an infinite population, evolutionary dynamics can result in either a Nash equilibrium of punishing and non-punishing cooperators or an oscillating state without punishers~\cite{brandt2006punishing}. If a single cooperator (either a non-punishing cooperator or a punisher) can participate in the game and a punisher can punish the non-punishing cooperator even in the absence of defectors, the evolutionary dynamics result in the stable coexistence of altruistic punishers and non-punishing cooperators~\cite{fowler2005altruistic}. In a finite population, with the assistance of loners, altruistic punishers can prevail and even dominate the whole population for most of the time when mutations are rare~\cite{hauert2007via}. If loners can escape punishment, altruistic punishment prevails even under the threat of anti-social punishment~\cite{garcia2012leaving}.

The concept of voluntary participation has recently been expanded with the introduction of a variant known as exiters~\cite{shen2021exit}. Exiters differ from risk-averse loners who receive a small but positive payoff by opting out and generate the same payoff for their opponent. In contrast, exiters simply exit the game, accepting a small but positive payoff without generating any payoff for their opponent, to avoid being exploited by defectors. This scenario is exemplified by scientific projects relying on collaboration. When two cooperative researchers work together as planned, project success is likely. However, if one researcher exhibits free-riding tendencies or abandons the project for a smaller immediate payoff, the remaining researcher's chances of success diminish. Despite their apparent similarity, the subtle difference between these two strategies leads to drastically different outcomes~\cite{shen2021exit,hauert2002volunteering}. Exiting not only allows exiters to avoid exploitation by defectors but also negatively impacts cooperators by generating nothing for their opponent. As a result, in one-shot games where reciprocity mechanisms are absent, the evolutionary outcome tends to be mutual exit. On the other hand, the impacts of loners on both cooperators and defectors are more balanced as they generate the same payoff for their opponent, enabling coexistence with cooperators and defectors through cyclic dominance in a one-shot game ~\cite{hauert2002replicator,szabo2002evolutionary}. In contrast, exiters allow cooperation to flourish only when direct, indirect, or network reciprocity is present~\cite{shen2021exit}.

Although research has explored the impact of the exit option on the evolution of cooperation, its role in the emergence of altruistic punishment remains unclear~\cite{shen2021exit}. Despite similarities between cooperation and altruistic punishment, recent studies suggest that the underlying mechanisms supporting behavior in these scenarios may differ significantly~\cite{molleman2019people,szolnoki2013correlation,burton2021decoupling}. These differences motivate our investigation into whether the exit option can facilitate the emergence of altruistic punishment, even though it does not allow for the emergence of cooperation in one-shot games. Our specific objectives are to determine the extent to which exiters contribute to explaining altruistic punishment and to evaluate whether the effectiveness of voluntary participation in solving the punishment conundrum remains robust in the presence of this variant of voluntary participation strategy. To address these questions, we introduce the concept of an exit option and incorporate altruistic punishment into a two-stage prisoner's dilemma game. Initially, we analyze well-mixed populations, considering both finite and infinite scenarios for the extended prisoner's dilemma game. Subsequently, we shift our focus to networked populations and make an intriguing observation: although the exit option does not provide significant benefits to altruistic punishment in well-mixed populations, it enables the existence of altruistic punishment in networked populations. Additionally, we identify multiple dynamical phenomena, such as cyclic dominance and a bistable state, that can be observed in networked populations.

\begin{table}
\caption{\label{t01} Payoff matrix for the weak prisoner's dilemma game with altruistic punishment and an exit option.}
\begin{ruledtabular}
\begin{tabular}{ccccc}
~   & $AP$ & $NC$ & $D$ & $E$      \\
\hline
$AP$ & $1$ & $1$ & $-\gamma$ & $0$ \\
$NC$ & $1$ & $1$ & $0$ & $ 0$ \\
$D$   & $b-\beta$   & $b$ & 0 & 0          \\
$E$ & $\epsilon$  & $\epsilon$ & $\epsilon$ & $\epsilon$ \\
\end{tabular}
\end{ruledtabular}
\begin{minipage}{0.48\textwidth}
\justify
The extended weak prisoner's dilemma game contains four competing action types: altruistic punishers who cooperate and punish defectors($AP$), non-punishing cooperators who cooperate but do not punish defectors ($NC$), defectors who free ride on the non-punishing cooperators and do not punish ($D$), and exiters who exit the game irrespective of whom they encounter ($E$). The first row indicates that when an altruistic punisher, $AP$, meets another altruistic punisher $AP$, non-punishing cooperator $NC$, defector $D$, or exiter $E$, they earn a payoff equal to 1, 1, $-\gamma$, or 0, respectively. When a non-punishing cooperator meets another altruistic punisher, non-punishing cooperator, defector, or exiter, they earn a payoff equal to 1, 1, 0, or 0, respectively. Analogously, when a defector meets an altruistic punisher, non-punishing cooperator, defector, or exiter, they earn a payoff equal to $b-\beta$, $b$, 0, or 0, respectively. Finally, exiters earn a payoff equal to $\epsilon\in \left[0,1\right)$, irrespective of whom they meet, and their opponent receives nothing.
\end{minipage}
\end{table}
\section*{Methods}
We studied the evolution of altruistic punishment in a two-stage prisoner's dilemma game by introducing two other action types, altruistic punishment and exit. In the first stage, each individual must make a choice simultaneously between cooperation ($C$), defection ($D$), and exit ($E$). In the second stage, cooperators decide whether to punish the defectors at a personal cost to themselves $\gamma$. To the defectors, this means an imposed fine $\beta$. This process results in four possible actions:

\begin{itemize}[topsep=0pt,parsep=0pt,itemsep=0pt]
\item $AP$, cooperate and punish defectors. Those who cooperate and punish are altruistic punishers because they punish free riders even at the expense of its own interests .
\item $NC$, cooperate but do not punish defectors. These non-punishing cooperators are also known as second-order free riders because by free-riding on punishment save the the cost of punishing the defectors.
\item $D$, defect but do not punish. These are also known as first-order free riders.
\item $E$, exit the game in favor of a small but positive payoff $\epsilon$ irrespective of whom they encounter. They do not participate in these two stages.
\end{itemize}

In a typical prisoner's dilemma game, mutual cooperation (defection) generates the reward (punishment) $R$ ($P$). If one player cooperates and the other defects, the cooperative player gets the sucker's payoff $S$, and the defected player obtains the temptation to defect $T$. For simplicity, we choose the weak prisoner's dilemma game as our base model by setting $R=1$, $P=S=0$, $T=b$. To make exiting less valuable than cooperating, and to ensure that the weak prisoner's dilemma game satisfied the payoff ranking of the strict prisoner's dilemma game, $T>R>P>S$ was used. Additional limits placed on the parameters were $1 \leq b<2$ and $\epsilon<1$. The described above is summarized in table \ref{t01}. In this study, we held constant the cost of punishment $\gamma$ and the fine for defectors at 0.1 and 0.3, respectively, while varying the temptation to defect $b$. This approach enabled us to investigate the influence of the exit option on altruistic punishment in scenarios where punishment initially promotes cooperation as well as those where it does not. By doing so, we gained a comprehensive understanding of the effects of the exit option on the evolution of altruistic punishment.
 
\subsection*{Finite population}
We first considered a finite and well-mixed population of $N$ individuals. Each individual adopted the Moran process~\cite{nowak2006evolutionary}, also known as frequently dependent process, to select their action. At each time step, a randomly selected player $i$ with fitness $f_i=e^{s\Pi_i}$ ($\Pi_i$ is the actual payoff of the individual $i$ obtained through their interaction) updates its action by imitating the action of player $j$ with fitness $f_j=e^{s\Pi_j}$ who is selected with a probability proportional to its fitness. Here, $s$ is the selection strength, the condition of $s\to 0$ corresponds to the weak selection and evolution proceeds as a neutral drift. We further assumed that with a small probability $\mu$, players randomly select their action from the rest of the possible actions, otherwise the imitation process is governed by the Moran process. When the mutation rate is small ($\mu \to 0$), the population will either eliminate the mutant or allow it to fully invade the population, resulting in homogeneity with one dominant actor most of the time.

Suppose that there are only two actions in the population, i.e., action $A$ and $B$, and these actions can be one of the four actions among the full action set \{$a,b,c,d$\}. Here, the symbols $a, b, c, d$ represent $AP, NC, D$ and $E$, respectively. In a finite population of size $N$ with $x$ $A$ and $y=N-x$ $B$ actions, the average payoff of $\Pi_{xy}$ and $\Pi_{yx}$ to players with $A$ and $B$ actions are the following:
\begin{equation}
\begin{array}{lr}
\Pi_{AB}=\frac{(x-1)P_{AA}+ (N-x)P_{AB}}{N-1}\\
\Pi_{BA}=\frac{xP_{BA}+ (N-x-1)P_{BB}}{N-1}
\end{array},
\label{e01}
\end{equation}
where $P_{AB}$ is the payoff obtained from the single encounter of actors $A$ and $B$, and so does payoffs $P_{AA}$, $P_{BA}$, and $P_{BB}$. This allows us to describe the evolutionary dynamics of the population in terms of a Markov Chain of size 4~\cite{fudenberg2006imitation,imhof2005evolutionary,santos2011co}. Given the above assumptions, the probability to change the number of $x$ individuals with action $A$ in a population of $y=N-x$ individuals with action $B$ by $\pm 1$, $T^{\pm}_{AB}$ is:
\begin{equation}\label{e02}
\begin{array}{lr}
T^{+}_{AB}=\frac{xf_i}{xf_i+yf_j}\frac{y}{N}\\
T^{-}_{AB}=\frac{yf_j}{xf_i+yf_j}\frac{x}{N}
\end{array},
\end{equation}
and hence the fixation probability $\rho_{AB}$ of a single mutant actor $A$ within a population of $N-1$ $B$ actors can be derived as~\cite{nowak2004emergence,traulsen2006stochastic}:
\begin{equation}\label{eq03}
\rho_{AB}=\frac{1}{\sum\limits_{k=0}^{N-1}\prod\limits_{x=1}^k\frac{T^{-}_{AB}}{T^{+}_{AB}}}=\frac{1}{\sum\limits_{k=0}^{N-1}\prod\limits_{x=1}^k\frac{e^{s\Pi_{BA}}}{e^{s\Pi_{AB}}}}.
\end{equation}
The fixation probabilities $\rho_{AB}$ define the transition probabilities of the a Markov process between four different homogeneous states of the population, with the following associated transition matrix:

\begin{equation}\label{eq04}
\bordermatrix{%
       & AP        & NC       &D     &   E \cr
AP    & \rho_{aa} &  \frac{\mu \rho_{ab}}{3}       & \frac{\mu \rho_{ac}}{3}      &  \frac{\mu \rho_{ad}}{3} \cr
NC    &  \frac{\mu \rho_{ba}}{3}         & \rho_{bb}       & \frac{\mu \rho_{bc}}{3}     &  \frac{\mu \rho_{bd}}{3}\cr
D     &  \frac{\mu \rho_{ca}}{3}          &  \frac{\mu \rho_{cb}}{3}      &\rho_{cc}      &  \frac{\mu \rho_{cd}}{3}\cr
E    &  \frac{\mu \rho_{da}}{3}          &  \frac{\mu \rho_{db}}{3}      & \frac{\mu \rho_{dc}}{3}     & \rho_{dd}
}.
\end{equation}
Here, $\rho_{AA}=1-\sum\limits_{A \neq B} \frac{\mu \rho_{AB}}{3}, A, B \in \{a,b,c,d\}$. The stationary distribution of each strategy, which is determined by the normalized right eigenvector to the largest eigenvalue, provides both the relative evolutionary advantage of each strategy and the stationary fraction of actors, as discussed in sources such as~\cite{sigmund2010social,hauert2007via}. For any pair of strategies $A$ and $B$ in the finite population, natural selection favors $B$ replacing $A$ only if $\rho_{AB}>\frac{1}{N}$~\cite{nowak2004emergence}.

\subsection*{Infinite population}
We then employed replicator dynamics to analyze the evolutionary outcomes in an infinite and well-mixed population. Let $x, y, z, w$ denote the fractions of altruistic punishers ($AP$), non-punishing cooperators ($NC$), defectors ($D$), and exiters ($E$) in the population. Where $0 \leq x, y, z, w \leq 1$, and $x+y+z+w=1$. The replicator equations are:
\begin{equation}
\begin{array}{l}
\dot{x} = x\lr{\Pi_{AP}-\overline{\Pi}}, \\
\dot{y} = y\lr{\Pi_{NC}-\overline{\Pi}}, \\
\dot{z} = z\lr{\Pi_D-\overline{\Pi}},  \\
\dot{w} = w\lr{\Pi_E-\overline{\Pi}}.
\end{array}
\label{eq05}
\end{equation}
The symbols $\Pi_{AP}$, $\Pi_{NC}$, $\Pi_D$, and $\Pi_E$ denote the average payoff of altruistic punishers, non-punishing cooperators, defectors, and exiters. Whereas $\overline{\Pi}=x\Pi_{AP}+y\Pi_{NC}+z\Pi_D+w\Pi_E$ is the average payoff of the whole population. According to the defined payoffs in table \ref{t01}, we obtained the following equation:
\begin{equation}
\begin{array}{l}
\Pi_{AP}  = x+y-z\gamma \\
\Pi_{NC}  = x+y \\
\Pi_D  = x(b-\beta)+yb \\
\Pi_E  = \epsilon
\end{array}.
\label{eq06}
\end{equation}
Using the constraint $w=1-x-y-z$, we obtained:
\begin{equation}
\left\{
\begin{array}{l}
\dot{x} = f\lr{x,y,z} \\= x\lrs{\lr{1-x}\lr{\Pi_{AP}-\Pi_E}-y\lr{\Pi_{NC}-\Pi_E}-z\lr{\Pi_{D}-\Pi_E}}\\
\dot{y} = g\lr{x,y,z} \\= y\lrs{\lr{1-y}\lr{\Pi_{NC}-\Pi_{E}}-x\lr{\Pi_{AP}-\Pi_{E}}-z\lr{\Pi_{D}-\Pi_{E}}} \\
\dot{z} = h\lr{x,y,z} \\= z\lrs{\lr{1-z}\lr{\Pi_{D}-\Pi_{E}}-y\lr{\Pi_{NC}-\Pi_{E}}-x\lr{\Pi_{AP}-\Pi_{E}}}
\end{array}
\right.
\label{eq07}
\end{equation}
For the detailed stability analysis of each equilibria, please refer to the Supplementary Information.

\subsection*{Networked population}
Different with well-mixed populations, global interactions in which an individual can interact with any other individual are no longer possible in the networked population. Instead, networks only allow local interactions, which means that individuals can only interact with their direct neighbors. Our basic network structure is a two dimensional regular lattice with periodic boundary conditions, each node was occupied by one individual, and each individual can only interact with its neighbors along its links. Our simulation contained the following steps. Initially, each individual was designed as either an altruistic punisher ($AP$), a non-punishing cooperator ($NC$), a defector ($D$), or an exiter ($E$) with equal probability. Each player acquires their total payoff by playing with all their direct neighbors according to the payoff matrix defined in table \ref{t01}. A randomly selected player $i$ decides to imitate the strategy of player $j$ who is also randomly selected from all the direct neighbors of player $i$ by comparing their payoff difference with the following probability:
\begin{equation}
W_{i\leftarrow{}j}=\frac{1}{1+\exp\left(\left(\Pi_i-\Pi_j\right)/K\right)},
\label{e08}
\end{equation}
where $\Pi_i$ and $\Pi_j$ is the acquired total payoff of the focal player $i$ and its randomly selected neighbor $j$, respectively. $K$ denotes the noise in the imitation process, and we fixed the value of $K$ to be 0.1 throughout the study.

A full Monte Carlo step is to repeat the above procedure $L^2$ times, and $L^2$ is the number of nodes in the given network. Each individual update their strategy once on average. To subside the transient dynamics and avoid the finite-size effect, we ran simulations for 50,000 steps on a regular lattice with size ranging from 200*200 to 800*800. The final fraction of each strategy was obtained after up to 45,000 steps. The presented data was averaged over 20 independent runs.

\section*{Results}
\subsection*{Well-mixed populations}
\begin{figure}
    \centering
    \includegraphics[scale=0.42]{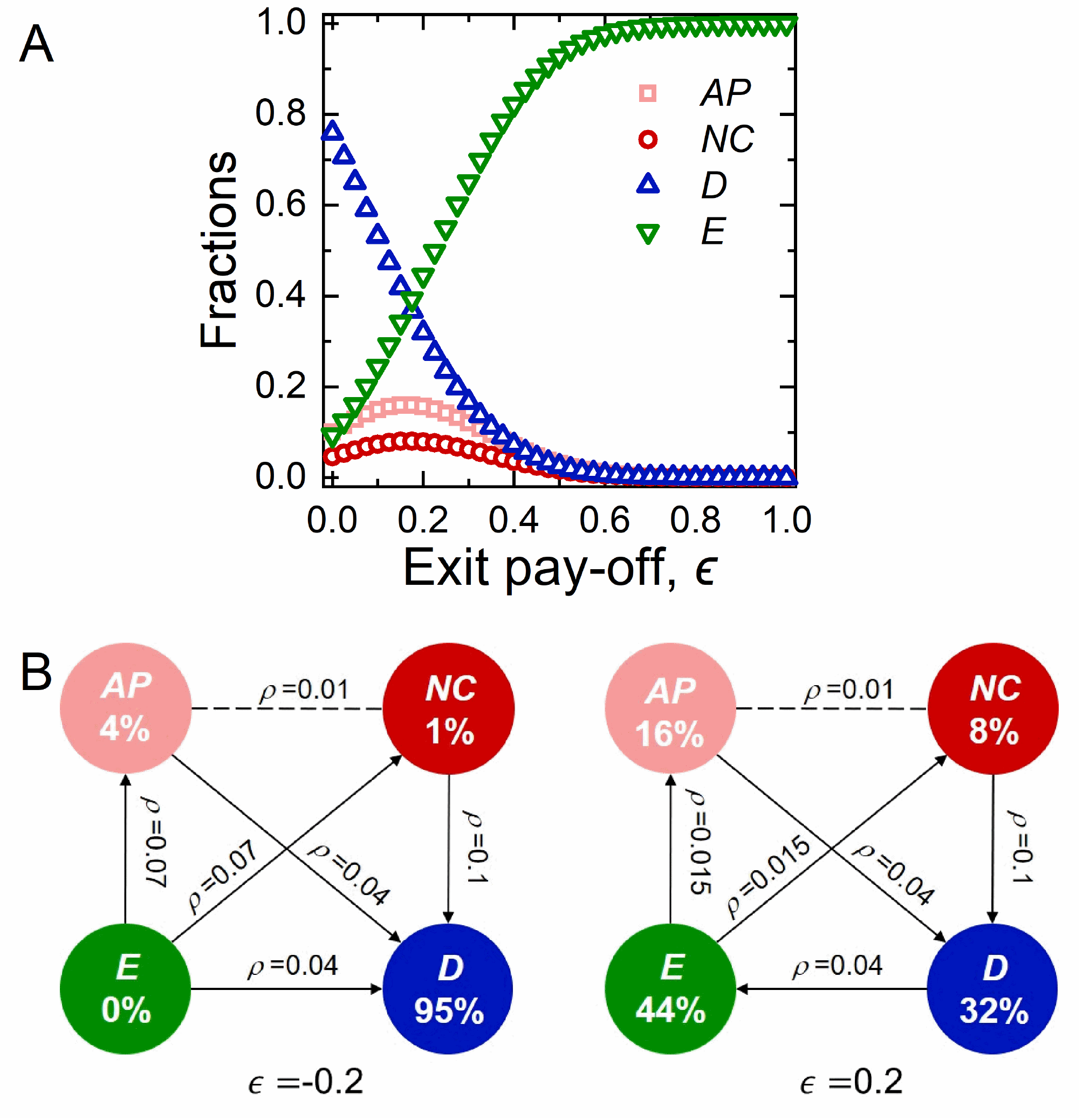}
    \caption{\textbf{Exiters establish altruistic punishment in a finite population, but altruistic punishers struggle to dominate the population.} A. Stationary probability distributions of each actors independence on the exiters' payoff $\epsilon$. B. Transition probabilities for each pair of actors when the exiters' payoff is negative (left) and positive (right). The parameter values are $b=1.5$, $\beta=0.3$, $\gamma=0.1$, $s=0.2$, $N=100$. }
    \label{f01A}
\end{figure}

\paragraph*{Finite population.}
In accordance with the existing literatures~\cite{egas2008economics,szolnoki2017second}, the maintenance of cooperation in the prisoner's dilemma game with altruistic punishment is contingent upon the cost and impact of punishment. If punishment is low-cost and highly impactful, cooperation can prevail. Conversely, if punishment is costly and less impactful, defectors will eventually dominate the entire population. In our extended model, we introduced negative payoffs for exiters, effectively transforming the game into a traditional weak prisoner's dilemma with altruistic punishment. In this scenario, the fine imposed on defectors remained larger than the payoff for cooperators, resulting in the dominance of defectors in the population (see left panel in Figure \ref{f01A}B and Figure \ref{fa1}A). However, when the payoff of exiters was slightly positive, altruistic punishers were able to coexist with defectors through cyclic dominance (see left panel in Figure \ref{f01A}B and Figure \ref{fa1}B). By increasing the payoff of exiters, denoted as $\epsilon$, we initially observed a peak fraction of altruistic punishers below 0.2, which subsequently declined until extinction (see Figure \ref{f01A}A). Importantly, the presence of exiters also enabled the existence of non-punishing cooperators through an alternative route of cyclic dominance, where $NC$ gave way to D, who gave way to $E$, who in turn gave way to $NC$ (refer to the right panel in Figure \ref{f01A}B), thus supporting the survival of non-punishing cooperators. In summary, while exiters facilitated the evolution of altruistic punishers in a finite population, they also allowed for the survival of second-order free riders, preventing the dominance of altruistic punishers in the entire population.

\paragraph*{Infinite population.}
Replacing the finite population with an infinite population significantly alters the evolutionary outcomes (theoretical analyses are presented in Appendix). Stability analysis yields the following results: (i) When $b-\beta>1$ and $\epsilon<0$, the monomorphic defecting equilibrium is stable, while the other equilibria are unstable (Figure \ref{fa2}A). (ii) When $b-\beta>1$ and $\epsilon>0$, the monomorphic exiting equilibrium is stable, and the other equilibria are unstable (Figure \ref{fa2}B). (iii) When $b-\beta<1$ and $\epsilon<0$, the evolutionary dynamics lead to either a mixed equilibrium of altruistic punishers and non-punishing cooperators or the monomorphic defecting equilibrium (Figure \ref{fa2}C). (iv) When $b-\beta<1$ and $\epsilon>0$, the evolutionary dynamics result in either a mixed equilibrium of altruistic punishers and non-punishing cooperators or the monomorphic exiting equilibrium (Figure \ref{fa2}D). In summary, in the infinite population, exiters only support the emergence of altruistic punishment when the cost-to-fine ratio of punishment is favorable for cooperators (i.e., $b-\beta>1$ in the model). Otherwise, exiters dominate the population. However, regardless of whether altruistic punishment can establish cooperation, the presence of exiters destabilizes defection and eventually replaces it.

In conclusion, our findings indicate that introducing the exit option in well-mixed populations provides little additional benefit for the dominance of altruistic punishment. Instead, the equilibrium is either monomorphic exiting in an infinite population or a joint dominance between defectors and exiters in a finite population. Unlike in well-mixed populations, cooperation can form compact clusters in networked populations to resist the invasion of defectors, a phenomenon known as network reciprocity~\cite{nowak1992evolutionary}. Such differences raise the question of what happens when the exit option is available in networked populations, where the survival of cooperation relies on compact cooperative clusters.

\subsection*{Networked population}
\begin{figure}[!t]
\includegraphics[scale=0.51]{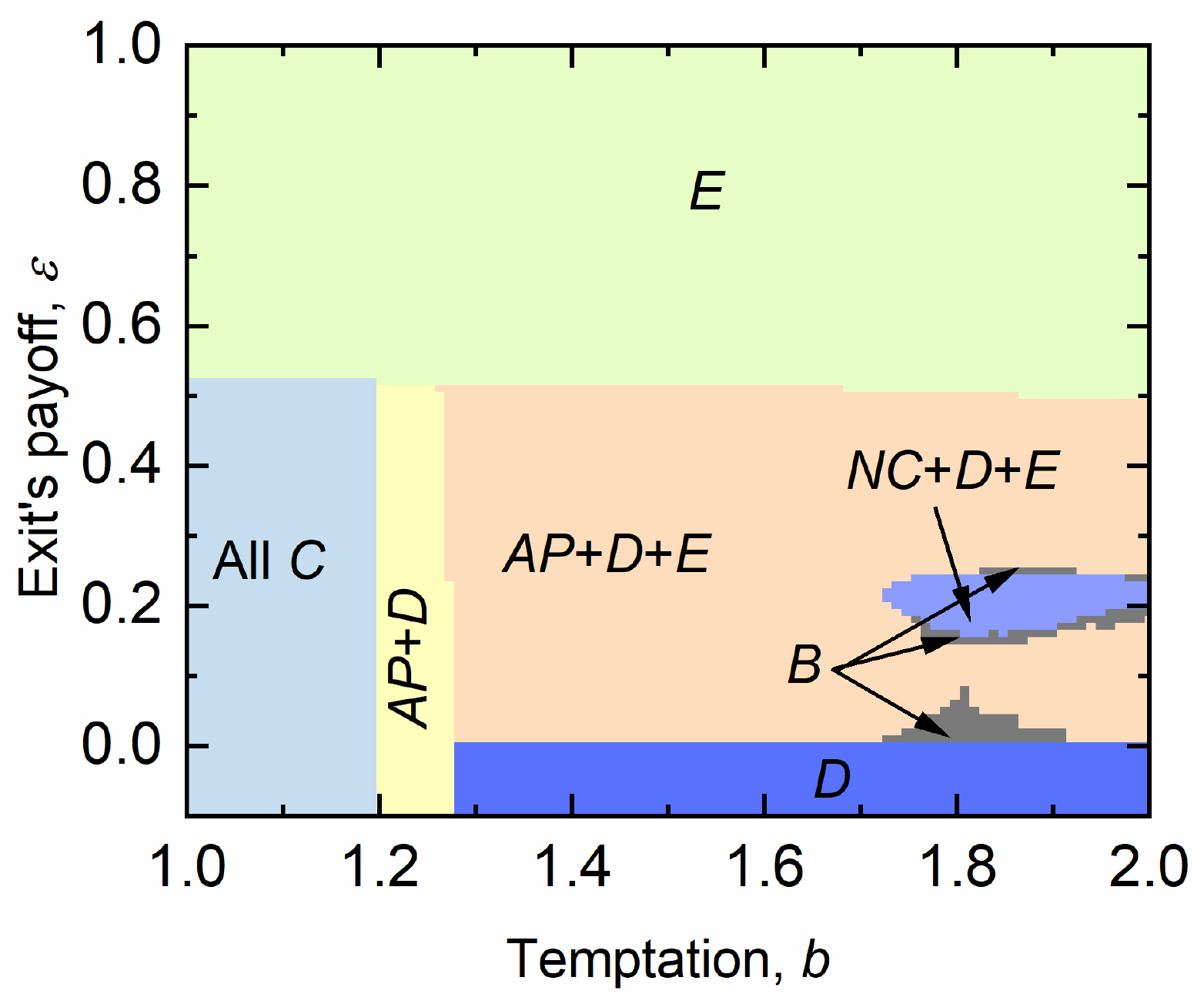}
\caption{\textbf{Exit option establishes altruistic punishment in networked population.} Presented is the full $\epsilon- b$ phase diagram obtained by Monte Carlo simulations of the extended weak prisoner's dilemma game on a regular lattice. Exiters dominate the whole population when the incentives to the exiters are large, $\epsilon \gtrsim 0.51$. Fewer exit option incentives lead to six possible outcomes. If $b$ is relatively small, $b\lesssim 1.19$, the effectiveness of altruistic punishment ensures the dominance of cooperators, and, altruistic punishers can coexist with defectors when $1.19 \lesssim b \lesssim 1.29$. For large temptation $b$, $b\gtrsim 1.29$, negative $\epsilon$ leads to full defection, whereas, positive $\epsilon$ ensures the coexistence of altruistic punishers with defectors and exiters, the coexistence of second-order free riders with defectors and exiters, or the bi-stable state of these two coexistence types.}
\label{f01}
\end{figure}

\begin{figure*}[!t]
\includegraphics[scale=0.51]{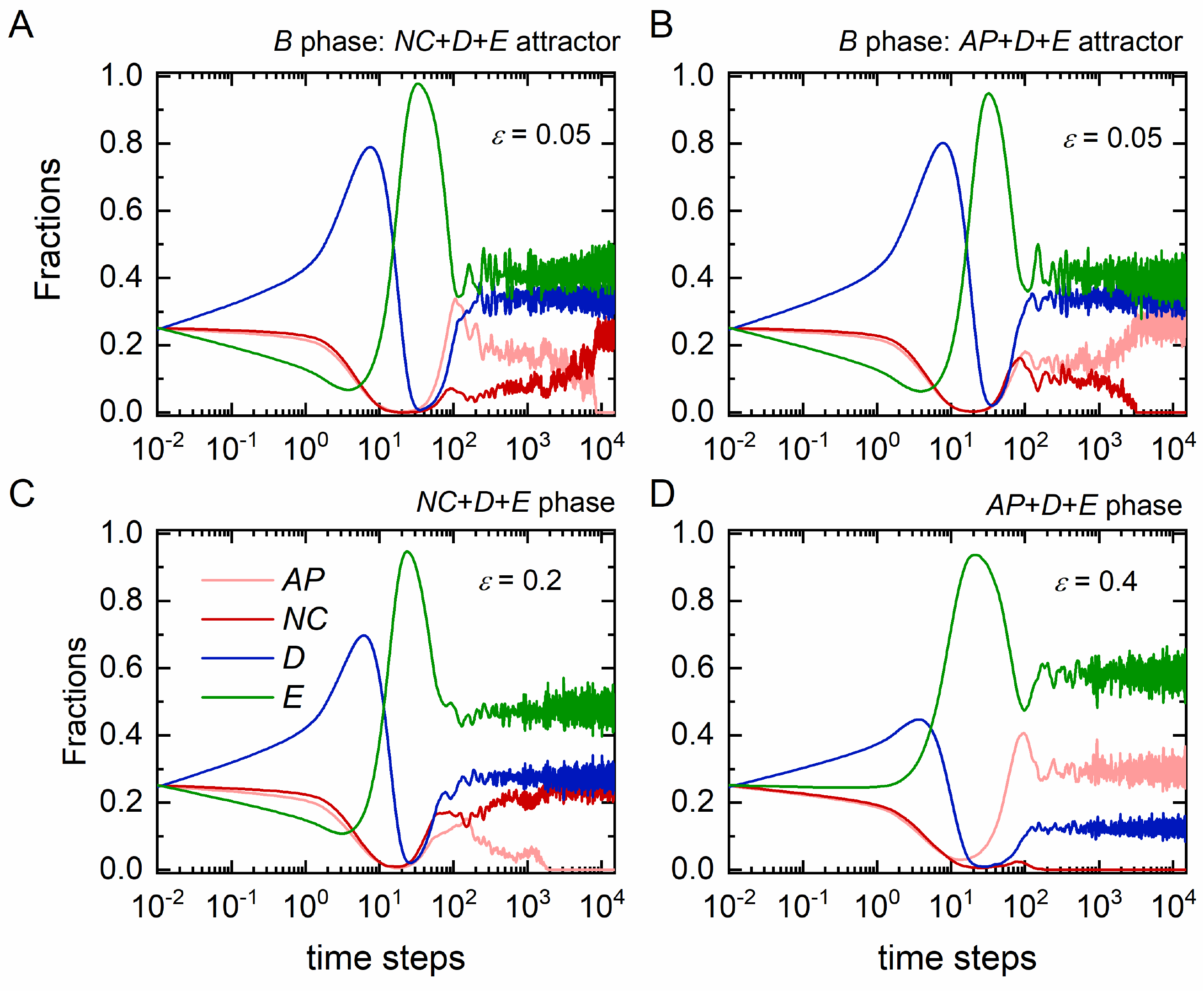}
\caption{\textbf{Time dependence of actor abundances exhibits complicated evolutionary dynamics.} In the bi-stable phase, starting from random initial conditions, small incentives to exit option lead the system to either $NC+D+E$ or $AP+D+E$ attractor but the coexistence of these four actors is not possible. During the evolution, if the abundance of altruistic punishers in the initial stage is much larger than that of the non-punishing cooperators, then altruistic punishers are eliminated and non-punishing cooperators coexist with defectors and exiters through cyclic dominance (figure.\ref{f02}A). However, if the abundance of altruistic punishers in the initial stage is comparable to that of non-punishing cooperators, then the non-punishing cooperators are eliminated and altruistic punishers coexist with defectors and exiters through cyclic dominance (figure.\ref{f02}B). Larger incentives to exiters turn the bi-stability to monostability and the evolutionary outcomes are determined by the incentives that were presented to exiters. The parameters were fixed as $b=1.8$, $\epsilon=0.05$ (top rows), $\epsilon=0.2$ (bottom left), and $\epsilon=0.4$(bottom right).}
\label{f02}
\end{figure*}

Figure.\ref{f01} shows the full $\epsilon-b$ phase diagram obtained by the extensive Monte Carlo simulations. It is noted that the addition of the simple exit option leads to complicated evolutionary outcomes. Initially, when the incentives to exiters are sufficiently large, $\epsilon \gtrsim 0.51$,  the exiters outcompete other action types and dominate the whole population (the $E$ phase in Figure.\ref{f01}), and this is consistent with previous findings~\cite{shen2021exit}. Less incentives to exiters, $\epsilon \lesssim 0.51$, lead to six different possible outcomes. In detail, if the temptation to defect is relatively small, $b\lesssim 1.29$, altruistic punishment together with network reciprocity are sufficient to maintain prosocial behavior (the All $C$ phase and the $AP+D$ phase in Figure.\ref{f01}). When $b\lesssim 1.19$, defectors can be completely eliminated by altruistic punishers, and thus altruistic punishers and non-punishing cooperators can coexist in a regular lattice. In the absence of defectors, non-punishing cooperators and altruistic punishers cannot be distinguished, and whether the evolutionary dynamics lead to the full $AP$ state, the full $NC$ state or the mixed $AP+NC$ state are determined by the initial conditions (the All $C$ phase in Figure.\ref{f01}). With increasing $b$, $ 1.19 \lesssim b \lesssim 1.29$, the effectiveness of altruistic punishment is greatly reduced, and defectors cannot be completely eliminated by altruistic punishers, and they coexist with the altruistic punishers in the population (the $AP+D$ phase in Figure.\ref{f01}). It is well established that altruistic punishment together with network reciprocity promotes cooperation even in the presence of antisocial punishment or second-order free-riders when the cost-to-fine ratio of punishment is low ~\cite{brandt2003punishment,helbing2010punish,szolnoki2017second}. The results of this study confirmed this conclusion. If $b$ is sufficiently large, altruistic punishment loses its effectiveness in sustaining prosocial behavior, and defectors dominate the entire population for negative $\epsilon$ (the $D$ phase in Figure.\ref{f01}). Non-negative values of exiters' payoff converted the full defection state to three possible outcomes: (i) the coexistence of $AP$, $D$ and $E$ (the $AP+D+E$ phase in Figure.\ref{f01}), (ii) the coexistence of $NC$, $D$, and $E$ (the $NC+D+E$ phase in Figure.\ref{f01}), or (iii) the bi-stable state between these two types of coexistences (the $B$ phase in Figure.\ref{f01}). In summary, when network reciprocity alone favors the existence of cooperation, the introduction of an exit option does not affect the evolutionary outcomes. However, when network reciprocity alone is insufficient to sustain cooperation (i.e., when $b>1.29$), exiters play a crucial role in maintaining cooperation within the networked population. They support cooperation by facilitating the coexistence of two distinct types of individuals: altruistic punishers and non-punishing cooperators. Interestingly, these two types of cooperators are unable to coexist within the networked population.

\begin{figure*}[!t]
\includegraphics[scale=0.51]{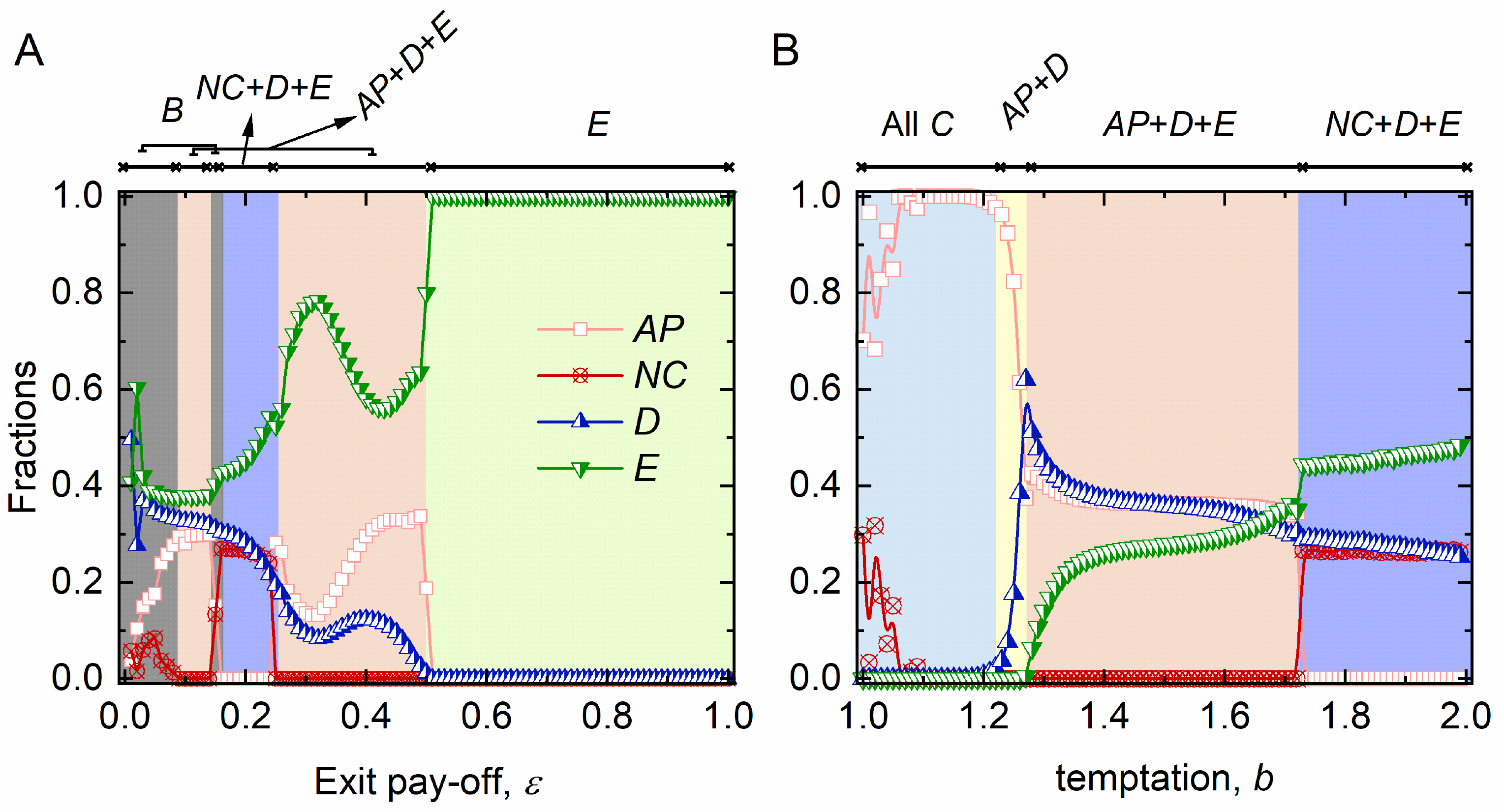}
\caption{\textbf{Power relations between altruistic punishers, second-order free riders, defectors, and exiters exhibiting complicated equilibra.} A. Along the vertical transect of $\epsilon-b$ phase at $b=1.8$. When $\epsilon \lesssim 0.06$, the networked population falls into the bi-stable state between the coexistence type of altruistic punishers, defectors, and exiters and the coexistence type of second-order free riders, defectors and exiters. In the range $0.06 \lesssim \epsilon \lesssim 0.16$, altruistic punishment outcompetes the second-order free riders, and coexists with the defectors and exiters. Whereas, in the range $0.16 \lesssim \epsilon \lesssim 0.17$, there is narrow dominance of the bi-stable state. In the range $0.17 \lesssim \epsilon \lesssim 0.25$, the second-order free riders outcompete the altruistic punishers, and coexist with the defectors and exiters. When $0.25 \lesssim \epsilon \lesssim 0.51$, the coexistence of altruistic punishers, defectors, and exiters again dominates the population. Finally the eixters dominate the population when $0.51 \lesssim \epsilon$. B. Along the horizontal transect of $\epsilon-b$ phase plane at $ \epsilon= 0.2$, the effectiveness of altruistic punishment together with network reciprocity is sufficient to secure prosocial behavior when $b \lesssim 1.29$. With increasing $b$, altruistic punishment loses its efficiency to sustain prosocial behavior, and adding exit option enables the networked population to first enter a coexistence state of altruistic punishers, defectors, and exiters in the temptation range of $1.29 \lesssim b \lesssim 1.73$, and reaches a coexistence state between second-order free riders, defectors, and exiters when $b \gtrsim 1.73$.}
\label{f03}
\end{figure*}

To gain a better understanding of how these actors coexist in the population, the evolution features of the fractions of each actors was examined and the results are presented in figure.\ref{f02}. In the bi-stable phase, it is the cooperators (altruistic punishers or non-punishing cooperators) start giving way to the defectors and with fewer cooperators around, defectors then giving way to the exiters. With large numbers of exiters, both the altruistic punishers and non-punishing cooperators compete for the exiters as they can only survive by adhering to the exiters. The described phenomenon is the cyclic dominance in which these actors dominate one another. Here, the cyclic dominance routes can be either (i) altruistic punishers that dominate exiters, who dominate defectors, who in turn dominate the altruistic punishers; or (ii) non-punishing cooperators that dominate the exiters, who dominate the defectors, who then dominate the non-punishing cooperators. As a key mechanism, researchers have verified the efficiency of cyclic dominance in sustaining bio-diversity or promoting cooperation~\cite{reichenbach2007mobility,szolnoki2014cyclic}. Although we started with random initial conditions, the evolutionary outcomes are different by implementing more independent simulations under same parameter combinations. For example, in the $NC+D+E$ attractor (figure.\ref{f02}A), the fraction of altruistic punishers is temporarily much larger than that of non-punishing cooperators at around 100th step, then the faction of altruistic punishers gradually decreases until it is eliminated and the fraction of second-order free riders increases until it reaches a stable state. However, in the $AP+D+E$ attractor (figure.\ref{f02}B), the fraction of altruistic punishers is always comparable to that of non-punishing cooperators up to around 1000th step, after this critical time step, the fraction of non-punishing cooperators gradually decreases until it is eliminated, and altruistic punishers gradually increase to reach a stable state. Thus, it is the initial distributions of the actors which determines the fate of altruistic punishers and non-punishing cooperators.
\begin{figure*}[!t]
\includegraphics[scale=0.22]{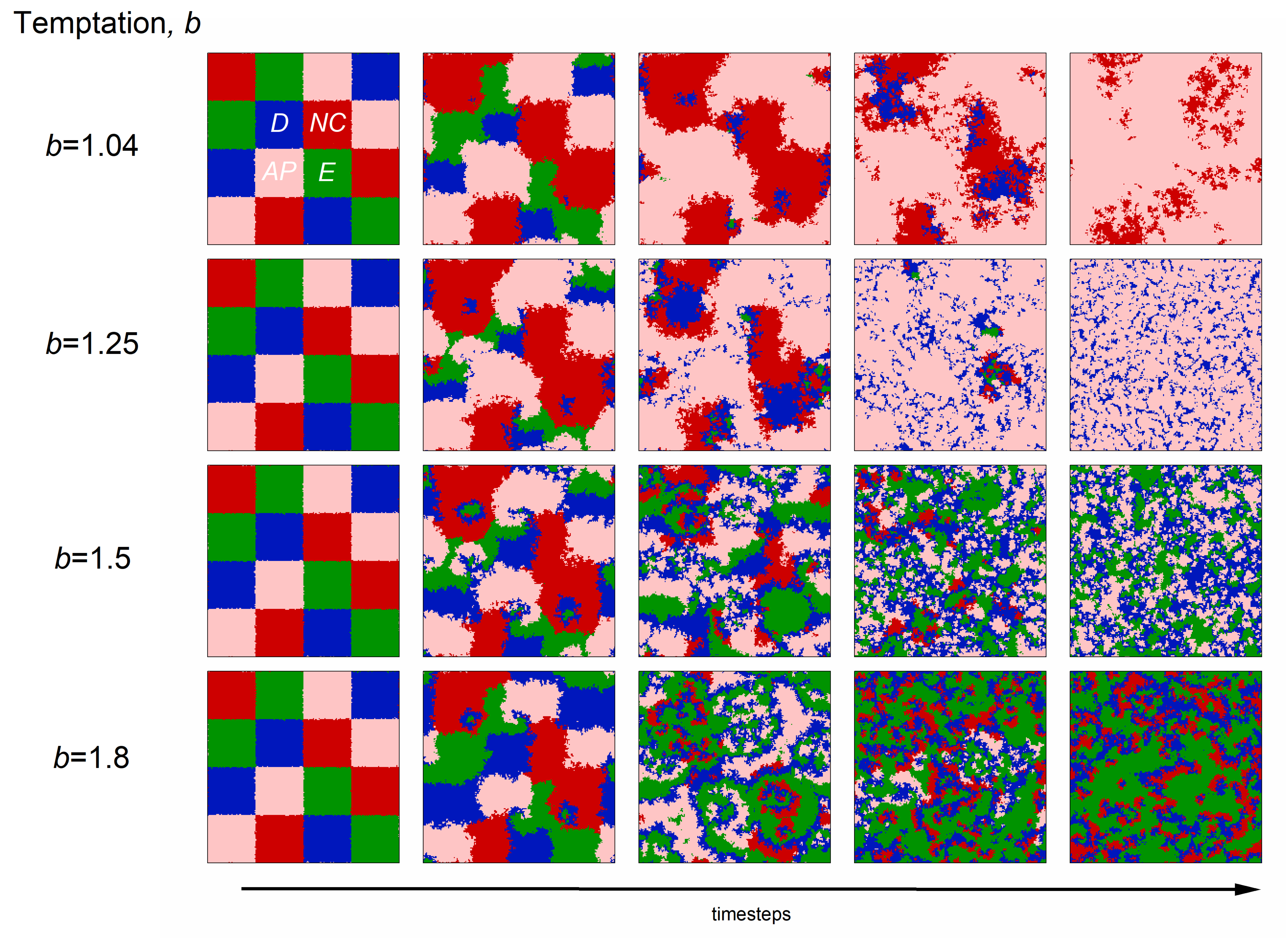}
\caption{\textbf{Evolutionary snapshots reveal the detailed dominance modes between all actors.} Shown are evolutionary snapshots at different time steps (column) and for different temptations for defection (rows). When the temptation is small (top row), both altruistic punishers and second-order free riders dominate the exiters, who take over the defectors. However, the decrease of exiters is much fast than its increase, and they are eliminated first. The defectors are then eliminated by the altruistic punishers, and finally the altruistic punishers coexist with second-order free riders in the population, and these two actors cannot separately be distinguished. When the temptation is larger (second row), the fate of exiters is the same as in the first row, however, the larger temptation leads more competitive defectors. Therefore, instead of completely dominating the defectors, the altruistic punishers coexist with defectors who replace the second-order free-riders until second-order free-riders they are eliminated. When the temptation is even larger (third row), more competitive defectors can encroach on both, the altruistic punishers and second-order free riders can only survive when they adhere to exiters. The indirect competition between altruistic punishers and second-order free riders with exiters determine the outcome for these two actors. Compared with non-punishing cooperators, altruistic punishers have greater fitness when compared to defectors and have greater probability to endure, therefore non-punishing cooperators are eliminated, and altruistic punishers coexist with defectors and exiters through cyclic dominance. When the temptation is at its largest (bottom row), exiters dominate and non-punishing cooperators have a larger probability to endure than altruistic punishers as it avoids the cost of punishment. Finally altruistic punishers are eliminated and non-punishing cooperators coexist with defectors and exiters. Results were obtained with $\epsilon = 0.2$ after the 30000th step to generate the final snapshots (rightmost column). The intermediate snapshots (second to fourth columns) were taken at different time steps across rows to ensure that the figure as illustrative as
possible.  }
\label{f04}
\end{figure*}
The phenomenon of bistability disappears when the incentives for exiters are increased. The evolutionary dynamics result in two distinct phases: the $NC+D+E$ phase and the $AP+D+E$ phase, depending on the incentives for exiters. When the fraction of exiters reaches its peak, both altruistic punishers and non-punishing cooperators can dominate over exiters. However, when the incentives for exiters are at an intermediate level ($\epsilon=0.2$), it is the non-punishing cooperators that dominate over exiters. Altruistic punishers lose in indirect competition with non-punishing cooperators and are eventually eliminated as the simulation proceeds. In this scenario, non-punishing cooperators coexist with defectors and exiters through cyclic dominance in the networked population (Figure \ref{f02}C). On the other hand, if the incentives for exiters are higher ($\epsilon=0.4$), altruistic punishers begin to dominate over exiters when exiters reach their first peak, and the non-punishing cooperators are unable to surpass the exiters and eventually get eliminated. In this case, altruistic punishers coexist with defectors and exiters through cyclic dominance in the system (Figure \ref{f02}D).

To understand the quantitative power relationships, i.e., which actors dominate each other under different conditions, we present two representative cross sections of the phase diagram in Figure \ref{f03}. Figure \ref{f03}A shows the stationary fractions of the four competing actors along the vertical transect of the $\epsilon-b$ phase plane, with $b=1.8$. In the traditional weak prisoner's dilemma game, where only cooperators and defectors exist, a high temptation leads to the complete dominance of defectors, rendering network reciprocity inefficient in supporting the coexistence of cooperators and defectors~\cite{szabo2005phase}. Although the inclusion of altruistic punishment in the weak prisoner's dilemma game can prevent this unfavorable outcome, its effectiveness in reducing defection comes at the expense of social welfare. The decrease in defection can only be achieved if the cost-to-fine ratio of altruistic punishment is relatively low, meaning a small punishment cost ($\gamma$) or a large punishment fine ($\beta$)~\cite{egas2008economics, szolnoki2017second, helbing2010evolutionary, helbing2010punish}. Conversely, if the cost-to-fine ratio of altruistic punishment is relatively large, altruistic punishment, along with network reciprocity, fails to provide sufficient benefits for cooperators, resulting in the continued dominance of defectors as the Nash equilibrium. However, introducing the exit option to the weak prisoner's dilemma game with altruistic punishment dramatically alters the equilibrium, even under unfavorable conditions that do not typically support cooperation for altruistic punishment.

When exit is costly ($\epsilon<0$), the defectors dominate the whole population (the $D$ phase in figure.\ref{f01}). As shown in figure.\ref{f03}A, if the incentives to exiters are small but positive, the $D$ phase gives way to the $B$ phase, where the system converges to either the $AP+D+E$ attractor or the $NC+D+E$ attractor depending on the results of the indirect competition between the altruistic punishers and non-punishing cooperators. By further increasing the $\epsilon$, the $NC+D+E$ phase is reached at $\epsilon \approx 0.17$, and there are two narrow strips that $AP+D+E$ phase and $B$ phase can dominate separately during this increment. The $AP+D+E$ phase dominates in the range $0.06 \lesssim \epsilon \lesssim 0.16$, and the $B$ phase is short-lived again in the range $0.16 \lesssim \epsilon \lesssim 0.17$. As $\epsilon$ continues to increase, the $NC+D+E$ phase gives way to $AP+D+E$ phase via discontinuous phase transition at $\epsilon \approx 0.25$. When incentives to exiters are sufficiently large, the $AP+D+E$ phase is finally replaced by the $E$ phase at the critical point $\epsilon \approx 0.51$.

Figure.\ref{f03}B shows the horizontal transect of $\epsilon-b$ at $\epsilon=0.2$, it also reveals the power relations between these competing actors, but it is dependent on the temptation level, $b$. When $b$ is small, $1 \leq b \lesssim 1.29$, the altruistic punishment together with the network reciprocity are able to support prosocial behavior. When $1 \leq b \lesssim 1.23$, the altruistic punishers can completely eliminate the defectors, the elimination of the defectors also negatively affects the exiters, and thus altruistic punishers coexist with non-punishing cooperators as they cannot be distinguished in the absence of defectors. The All-$C$ phase transitions to the $AP+D$ phase through a continuous phase transition. Although the advantages of cooperators decrease with increasing $b$, those cooperators who punish defectors gain a greater advantage compared to defectors. Therefore, in this scenario, network reciprocity supports the coexistence of altruistic punishers and defectors. If the conditions to support cooperation with altruistic punishment are unfavorable, adding an exit option can promote the system to the $AP+D+E$ phase when $b \lesssim 1.73$. However, with increasing $b$, the $AP+D+E$ phase gives way to the $NC+D+E$ phase through discontinuous phase transition at the critical point, $b \approx 1.73$.

\begin{figure}[!t]
\includegraphics[scale=0.16]{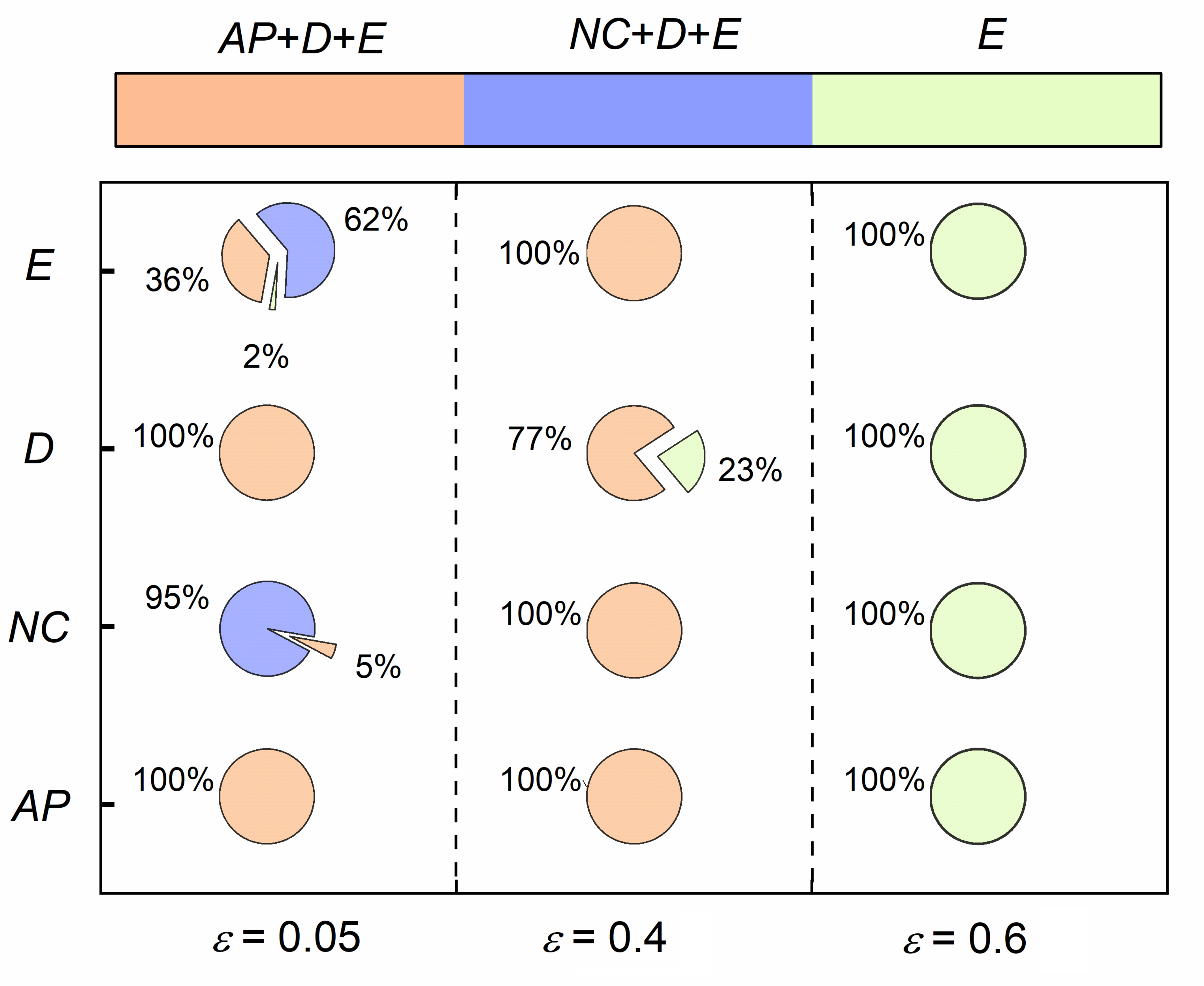}
\caption{\textbf{Initial conditions determine the outcome of altruistic punishers and non-punishing cooperators in the bi-stable phase.} Shown are the evolutionary outcomes after implementing 100 independent simulations for each parameter combination under four different initial conditions. The initial conditions were (i) $97\%$ of players were initially assigned as $AP$, (ii) $97\%$ of players were initially assigned as $NC$, (iii) $97\%$ of players were initially assigned as $D$, and (iv) $97\%$ of players were initially assigned as $E$. The rest of the other action types were assigned to the other players with equal probability in these different initial conditions. Parameters were fixed as $b=1.8$, from left to right, $\epsilon =0.05, 0.4, 0.6$, respectively.}
\label{f05}
\end{figure}

To reexamine the evolutionary dynamics and further check the indirect competition between altruistic punishers and non-punishing cooperators in both spatial and temporal dimensions,. We plotted the evolutionary snapshots for varying $b$ at $\epsilon=0.2$, and the results are presented in figure.\ref{f04}. When the temptation is small (top row in figure.\ref{f04}), the exiters were eliminated first by altruistic punishers and non-punishing cooperators, and the defectors experienced the same fate shortly after. The altruistic punishers coexist with non-punishing cooperators eventually as they cannot be distinguished and the system falls into frozen state. A larger temptation makes the defectors more competitive (second row in figure.\ref{f04}), and instead of being eliminated by the altruistic punishers, they can coexist. However, the coexistence of defectors cannot ensure the survival of exiters, who are eliminated in situations with small temptation. The non-punishing cooperators are eliminated by defectors and finally, the altruistic punishers coexist with defectors in the population. When the temptation is even larger, the coexistence of defectors and altruistic punishers was no longer possible, instead, defectors can invade both altruistic punishers and non-punishing cooperators. The competitive defectors allow for the survival of exiters. In turn, altruistic punishers and non-punishing cooperators can survive by adhering to the survived exiters. It is therefore, both altruistic punishers and non-punishing cooperators can coexist with defectors and exiters through different cyclic dominance routes. However, these two types of cyclic dominance cannot coexist in the population, and the indirect competition to the territories of exiters between altruistic punishers and non-punishing cooperators determine the outcome of the competitors. Competitive defectors more easily negatively affexct non-punishing cooperators than altruistic punishers (third row and second column in figure.\ref{f04}), and therefore, non-punishing cooperators are eliminated first, and the altruistic punishers, defectors, and exiters coexist within the population. When the temptation is the largest (bottom row in figure.\ref{f04}), defectors are the most competitive, altruistic punishers and non-punishing cooperators are exploited by defectors at almost the same speed, and the exiters dominate by eliminating the defectors. In the indirect competition of exiters with the non-punishing cooperators, the altruistic punishers loses its advantages due to the existence of punishment cost, and non-punishing cooperators coexist with defectors and exiters.

Our results have shown the consideration of exit option results in the bi-stable dynamics and it is the initial distribution of actors determines the outcome of altruistic punishers and non-punishing cooperators. It is generally accepted that the initial conditions are crucial for evolutionary outcomes in agent-based models~\cite{perc2017statistical}. We further assessed whether the initial fractions of actors is a potential reason that the system exhibits bi-stability. Figure.\ref{f05} presents the evolutionary outcomes with $\epsilon=0.05, 0.4$, and $0.6$ under four different initial conditions. The four different conditions are: (i) $97\%$ of players were initially assigned as $AP$, (ii) $97\%$ of players were initially assigned as $NC$, (iii) $97\%$ of players were initially assigned as $D$, and (iv) $97\%$ of players were initially assigned as $E$. The other players were assigned one of the other three actions with equal probability in these conditions. The results were obtained by implementing 100 independent simulations. We found that when $\epsilon=0.05$ (left column in figure.\ref{f05}), the evolutionary outcome was always $AP+D+E$ if the majority of players initially had action $AP$ or action $D$. However, if the majority of players initially had $NC$ action, then the system reached the attractor $NC+D+E$ with $95\%$ probability. If the majority of players were $E$, then the system reached the attractor $AP+D+E$ or $NC+D+E$ with $36\%$ and $62\%$ probability, respectively. Larger incentives to exiters switched the bi-stability to monostability (middle and right column in figure.\ref{f05}). In the monostability state, evolutionary dynamics lead to either the $AP+D+E$ or the $E$ phase depending on the incentives to the exiters, and evolutionary outcomes are independent on the initial conditions. The finite-size effects are a potential pitfall that may generate misleading results when implementing agent-based models in structured populations~\cite{perc2017statistical}. Thus, it is crucial to choose a sufficiently large network size or to employ the method of subsystem solutions to avoid this potential issue~\cite{szolnoki2016competition,perc2017stability}. It is noteworthy that the system has $23\%$ probability to fall into the full $E$ phase when most players initially had $D$ action at $\epsilon=0.4$ (middle column in figure.\ref{f05}). We do believe that the counterintuitive $E$ phase is the product of the finite-size effect, and the pure $AP+D+E$ phase can be expected as long as a larger network size was implemented.

\section*{Discussion}
In our discussion, we have demonstrated that the introduction of an exit option in the two-stage prisoner's dilemma game leads to complex dynamics. Specifically, in an infinite and well-mixed population, we observed that exiters do not have any impact on the emergence of cooperation. Instead, they simply destabilize defection and eventually replace it, regardless of whether altruistic punishment alone can sustain cooperation. However, in a finite and well-mixed population, the availability of exit options ensures the survival of both altruistic punishers and second-order free riders through two types of cyclic dominance. Nevertheless, altruistic punishers never dominate the population in this scenario.

In contrast to well-mixed populations, when combining the exit option with network reciprocity, we discovered significantly different outcomes. We found that the dominance of altruistic punishment is possible in a networked population. Altruistic punishers can coexist with defectors and exiters through cyclic dominance in a majority of the $\epsilon-b$ phase plane. Notably, when the temptation is high ($b \gtrsim 1.71$), exiters facilitate the survival of second-order free riders. Furthermore, depending on the incentives provided to exiters, the system can also enter a bistable phase or a single $NC+D+E$ phase. Thus, it is evident that the exit option is not a panacea for solving the second-order free riding problems.

Exiters produce outcomes that differ greatly from these in loners. In the infinite and well-mixed population, adding an exit option can also result in a bi-stable outcome, in which the Nash equilibrium can be either the coexistence of altruistic punishers and non-punishing cooperators or a monomorphic exiting equilibrium. However, this bi-stable outcome is only possible when the punishment itself is sufficient to maintain cooperation, otherwise, the bi-stable outcome can be replaced with a monomorphic exiting equilibrium. In other words, exiters just simply destabilize the defectors and eventually replaces them in the infinite population. In the finite population, although exiters allow the survival of altruistic punishment when the exiter's payoff is moderate, altruistic punishers never dominate the whole population (e.g. figure.\ref{f01A}A). The direct comparison between exiters and loners in a finite and infinite population lead us to conclude that loners are more effective than exiters in supporting the prevalence of altruistic punishment.

The effectiveness of altruistic punishment is not only challenged by second-order free riders but also by the presence of antisocial punishment, which has been observed in various human cultures through experimental studies~\cite{herrmann2008antisocial,denant2007punishment,gachter2011limits}. Recent theoretical studies have demonstrated that the existence of antisocial punishment can impede the successful coevolution of punishment and cooperation~\cite{rand2009direct,rand2010anti}. Additionally, when punishment is accessible to loners, it does not promote cooperation but instead becomes a self-serving tool for protecting oneself against potential competitors~\cite{rand2011evolution}. As mentioned earlier, exiters can be considered spiteful punishers as they harm both cooperators and defectors, while loners generate a small yet positive payoff for their opponents. This slight disparity leads to significantly different equilibrium outcomes in a one-shot game, where exiters can destabilize defectors and eventually replace them, while loners can sustain cooperation through cyclic dominance. To comprehensively explore the stability of punishment and its implications, it is essential to consider all possible punishment sets, including antisocial and spiteful punishment, and allow actors to punish each other. Such an extension to our model can shed light on how this setup influences the stability of punishment and whether it yields different outcomes compared to those observed in loners. Therefore, these findings warrant further investigation.

Further verification is crucial to determine whether exit options can effectively foster altruistic punishment and eliminate second-order free riders through human behavior experiments. Experimental studies often produce contrasting or surprising results when compared to theoretical predictions. In this regard, a recent relevant study explored the introduction of punishment into networks as a means to theoretically promote cooperation \cite{brandt2003punishment, szolnoki2017second, helbing2010punish, perc2012self}. However, a large-scale human behavior experiment concluded that the introduction of peer punishment did not promote cooperation in structured populations, but instead reduced the benefits of network reciprocity \cite{li2018punishment}. While our demonstration supports the idea that exiters can facilitate the prevalence of altruistic punishment within the framework of network reciprocity, it remains imperative to design experimental models that can further validate our theory through human behavior experiments.

\section*{Article information}
\paragraph*{Acknowledgements.}  We thank Prof. Dr. Marko Jusup for valuable discussions. This research was supported by the National Science Fund for Distinguished Young Scholars (grants no. 62025602). We also acknowledge support from (i) a JSPS Postdoctoral Fellowship Program for Foreign Researchers (grant no. P21374), and an accompanying Grant-in-Aid for Scientific Research from JSPS KAKENHI (grant no. JP 22F31374), and the National Natural Science Foundation of China (grant no.~11931015) to C.\,S. as a co-investigator, (ii) the National Natural Science Foundation of China (grants no.~11931015, ~12271471 and 11671348) to L.\,S., (iii) National Natural Science Foundation of China (grants no. U22B2036, 11931015), Key Technology Research and Development Program of Science and Technology-Scientific
and Technological Innovation Team of Shaanxi Province (Grant No. 2020TD-013) and the XPLORER PRIZE. to Z.\,W, and (iv) the grant-in-Aid for Scientific Research from JSPS, Japan, KAKENHI (grant No. JP 20H02314) awarded to J.\,T.
\paragraph*{Author contributions.} C.\,S. and L.\,S. conceived research. C.\,S. and Z.\,S. performed simulations. All co-authors discussed the results and wrote the manuscript.
\paragraph*{Conflict of interest.} Authors declare no conflict of interest.

\bibliographystyle{elsarticle-num}
\bibliography{biblio}
\vspace{5mm}


\setcounter{equation}{0}
\renewcommand\theequation{S\arabic{equation}}
\renewcommand\thefigure{S\arabic{figure}}
\setcounter{figure}{0}
\clearpage
\onecolumngrid
\setcounter{equation}{0}
\renewcommand\theequation{S\arabic{equation}}
\setcounter{page}{1}
\renewcommand\thepage{S\arabic{page}}
\section*{
Supplementary Information for\\
``Exit options sustain altruistic punishment and decrease the second-order free-riders, but it is not a panacea''}

\section*{Stability analysis of the equilibria in infinite and well-mixed population}
Solving Eq.\ref{eq07}, we obtain 12 equilibrium points: $(1,0,0,0)$, 
$(0,1,0,0)$, 
$(0,0,0,1)$, 
$(0,0,1,0)$, 
$(\epsilon, 0,0,1-\epsilon)$,
$(0,\epsilon,0,1-\epsilon)$, 
$(x,1-x,0,0)$, $(x,\epsilon-x,0,1-\epsilon)$, 
$(\frac{-1+b}{\beta},\frac{1-b+\beta}{\beta},0,0)$,
$(\frac{\epsilon}{b-\beta},0,\frac{\epsilon-\beta-\epsilon\beta}{(\beta\epsilon)\gamma},1-\frac{\epsilon(\gamma+1+\beta-b)}{(b-\beta)\gamma})$,
$(\frac{(-1+b)\epsilon}{\beta},\frac{\epsilon-\beta+\beta\epsilon}{\beta},0,1-\epsilon)$,
$(\frac{\gamma}{1-b+\beta+\gamma},0,\frac{-1+b-\beta}{-1+b-\beta+\gamma},0)$.
To examine the stability of these equilibria, we calculate the eigenvalues of Jacobian matrix:
\begin{equation}
J=\left[
    \begin{array}{ccc}
       \frac{\partial f(x,y,z)}{\partial x}  &\frac{\partial f(x,y,z)}{\partial y} &\frac{\partial f(x,y,z)}{\partial z} \\
         \frac{\partial g(x,y,z)}{\partial x}  &\frac{\partial g(x,y,z)}{\partial y} &\frac{\partial g(x,y,z)}{\partial z} \\
         \frac{\partial h(x,y,z)}{\partial x}  &\frac{\partial h(x,y,z)}{\partial y} &\frac{\partial h(x,y,z)}{\partial z} 
    \end{array}
    \right].
\end{equation}

If the eigenvalues have negative real parts, Eq.\ref{eq07} will approach zero regardless of the initial states. Thus, when all eigenvalues have negative real parts, the corresponding equilibrium is stable. When some eigenvalues have positive real parts, the corresponding equilibrium is unstable. If some eigenvalues have negative real parts and the rest eigenvalues have zero real parts, the stability of equilibrium needs to be determined by the center manifold theorem. The stability can be determined by analyzing a lower-order system whose order equals the number of eigenvalues with zero real parts.

Then we have the following conclusion.

\begin{theorem}
When $b<1+\beta$, and $\epsilon<0$, the equilibrium points $(x^*,1-x^*,0,0)$ and $(0,0,1,0)$ are stable, while the rest of others are unstable; When  $b<1+\beta$, and $\epsilon>0$, the equilibrium points $(x^*,1-x^*,0,0)$ and $(0,0,0,1)$ are stable, and the others are unstable; When $b\ge1+\beta$, only the equilibrium point $(0,0,1,0)$ is stable, and the rest of others are unstable. When $\epsilon>0$, only the equilibrium point $(0,0,0,1)$ is stable, and the rest of others are unstable.
\end{theorem}

\begin{proof}
    (1). For $K_1$:  $(x,y,z,w)=(1,0,0,0)$, the Jacobian matrix $J_1$ is 
\begin{equation}
    J_1=\left[
    \begin{array}{ccc}
        -1+\epsilon & -1+\epsilon & -b+\beta+\epsilon \\
        0 & 0 & 0 \\
        0 & 0 & -1+b-\beta 
    \end{array}
    \right]
\end{equation}
and its corresponding eigenvalues are 
\begin{equation}
    \{\lambda_1,\lambda_2,\lambda_3\}=\{0,-1+b-\beta,-1+\epsilon\} .
\end{equation}
When $b>\beta+1$, $K_1$ is unstable because $-1+b-\beta$ is a positive eigenvalue. Otherwise, there is at least one zero eigenvalue. Thus, we use the center manifold theorem to analyze the stability of $K_1$. Using $b<\beta +1$ as an example. First, there is an invertible matrix whose column elements are the eigenvectors of $J_1$
\begin{equation}
    P=\left[
    \begin{array}{ccc}
         -1 & -1 & 1  \\
         1 & 0 & 0  \\
         0 & 1 & 0 
    \end{array}
    \right]
\end{equation}
and $J_1$ can be diagonalized as
 \begin{equation}
P^{-1}J_1P=\left[
    \begin{array}{ccc}
       0 & 0 & 0 \\
       0 & -1+b-\beta & 0 \\
       0 & 0 & -1+\epsilon
    \end{array}
    \right].
 \end{equation}
Then change of variable:
\begin{equation}
    \left[ 
    \begin{array}{c}
         x'_1  \\
         y_1  \\
         z_1
    \end{array}
    \right] = P^{-1} \left[
    \begin{array}{c}
         x  \\
         y  \\
         z
    \end{array}
    \right] = \left[
    \begin{array}{c}
         y  \\
         z  \\
         x+y+z
    \end{array}
    \right]
\end{equation}
and the system becomes Eq.(\ref{aaa1}).
\begin{figure*}
\begin{equation}
\centering
    \begin{split}
        \dot{x}'_1= &g(z_1-x'_1-y_1,x'_1,y_1)\\
        =&x'_1 ((1 - x'_1) (-\epsilon - y_1 + z_1)- x'_1 (-\epsilon + b x'_1 + (b - \beta) (-x'_1 - y_1 + z_1))-  y_1 (-\epsilon + b x'_1 + (b - \beta) (-x'_1 - y_1 + z_1))) \\
        \dot{y_1}= &h(z_1-x'_1-y_1,x'_1,y_1)\\
        =&y_1 (-x'_1 (-\epsilon - y_1 + z_1)-y_1 (-\epsilon - y_1 - x'_1 y_1 + z_1)+ (1 - y_1) (-\epsilon + b x'_1 + (b - \beta) (-x'_1 - y_1 + z_1)))\\
        \dot{z_1}= &f(z_1-x'_1-y_1,x'_1,y_1)+g(z_1-x'_1-y_1,x'_1,y_1) +h(z_1-x'_1-y_1,x'_1,y_1) \\
                = &x'_1 (\epsilon (-1 + 2 x'_1 + y_1) + (-1 + x'_1 + b x'_1 + b y_1) (y_1 - z_1) - \beta (x'_1 + y_1) (x'_1 + y_1 - z_1)) + \\
                &y_1 ((-1 + y_1) (\epsilon + b (y_1 - z_1) - \beta (x'_1 + y_1 - z_1)) + x'_1 (\epsilon + y_1 - z_1) + y_1 (\epsilon + y_1 + x'_1 y_1 - z_1)) +\\
                &(x'_1 + y_1 - z_1) (\epsilon + (1 - b + \beta + x'_1) y_1^2 -\epsilon z_1 + (-1 + z_1) z_1 + y_1 (1 + x'^{2}_1 + x'_1 (1 + \beta - z_1) +(-2 + b - \beta) z_1)).
    \end{split}
\label{aaa1}
\end{equation} 
\end{figure*}
Let $x'_1=x_1+1$ and the system becomes Eq.(\ref{aaa2}).
\begin{figure*}
\begin{equation}
    \begin{split}
        \dot{x_1}= &g(z_1-x_1-1-y_1,x_1+1,y_1)\\
        =&(x_1+1) (- x_1 (-\epsilon - y_1 + z_1)- (x_1+1) (-\epsilon + b (x_1+1) + 
        (b - \beta) (-x_1-1 - y_1 + z_1))- \\
        & y_1 (-\epsilon + b (x_1+1) + (b - \beta) (-x_1-1 - y_1 + z_1))) \\
        \dot{y_1}= &h(z_1-x_1-1-y_1,x_1+1,y_1)\\
        =&y_1 (-(x_1+1) (-\epsilon - y_1 + z_1)-
        y_1 (-\epsilon - y_1 - (x_1+1) y_1 + z_1)+ 
        (1 - y_1) (-\epsilon + b (x_1+1) + \\
        &(b - \beta) (-x_1-1 - y_1 + z_1)))\\
        \dot{z_1}= &f(z_1-x_1-1-y_1,x_1+1,y_1)+
        g(z_1-x_1-1-y_1,x_1+1,y_1) 
        +h(z_1-x_1-1-y_1,x_1+1,y_1) \\
        = &(x_1+1) (\epsilon (-1 + 2 (x_1+1) + y_1) + 
        (x_1 + b (x_1+1) + b y_1) (y_1 - z_1) - 
        \beta (x_1 + 1 + y_1) (x_1 + 1 + y_1 - z_1)) + \\
        & y_1 ((-1 + y_1) (\epsilon + b (y_1 - z_1) - \beta (x_1 + 1 + y_1 - z_1)) + 
        (x_1+1) (\epsilon + y_1 - z_1) + y_1 (\epsilon + y_1 + (x_1+1) y_1 - z_1)) + \\
        &(x_1 + 1 + y_1 - z_1) (\epsilon + (2 - b + \beta + x_1) y_1^2 -\epsilon z_1 + (-1 + z_1) z_1 + y_1 (1 + (x_1+1)^2 + 
        (x_1+1) (1 + \beta - z_1) +\\
        &(-2 + b - \beta) z_1)).
    \end{split}
\label{aaa2}
\end{equation}
\end{figure*}

Put the system into the form
\begin{equation}
    \begin{array}{c}
         \dot{\boldsymbol{X}}=A\boldsymbol{X}+\textbf{F}(\boldsymbol{X}, \boldsymbol{Y})  \\
         \dot{\boldsymbol{Y}}=B\boldsymbol{Y}+\boldsymbol{G}(\boldsymbol{X}, \boldsymbol{Y}) 
    \end{array},
\end{equation}
where $\boldsymbol{X}=[x_1]$, $\boldsymbol{Y}=\left[\begin{array}{c}
     y_1  \\
     z_1 
\end{array} \right]$, and $A=[0]$, $B=\left[\begin{array}{cc}
     -1+b-\beta & 0  \\
     0 & -1+\epsilon 
\end{array}\right]$, whose eigenvalues have zero and negative real parts, respectively. $\boldsymbol{F}$ and $\boldsymbol{G}$ are the functions of $\boldsymbol{X}$ and $\boldsymbol{Y}$. They satisfy the condition $\boldsymbol{F(0,0)=0}, \boldsymbol{F'(0,0)=O}$. According to the existence theorem of the center manifold, the system has the center manifold $S=\{(\boldsymbol{X}, \boldsymbol{H(X)})|\boldsymbol{H}:\mathbb{R}^1\to \mathbb{R}^2\}$. We define a mapping 
\begin{equation}
\begin{split}
        (M\varphi)(\boldsymbol{X})= &\varphi'(\boldsymbol{X})(A\boldsymbol{X}+\boldsymbol{F}(\boldsymbol{X},\varphi(\boldsymbol{X})) \\
        &-B\varphi(\boldsymbol{X})-\boldsymbol{G}(\boldsymbol{X},\boldsymbol{\varphi}(\boldsymbol{X}))
\end{split}
\end{equation}
Set $\varphi(\boldsymbol{Y})=O(\boldsymbol{X}^2)$, we obtain
\begin{equation}
\begin{split}
    \dot{x_1}=&(x_1+1) (-\epsilon x_1-(x_1+1) (-\epsilon + b (x_1+1) \\
    &-(x_1+1)(b - \beta)))+O(x_1^4)
\end{split}
\end{equation}
Then we define $m(x_1)=(x_1+1) (-\epsilon x_1-(x_1+1) (-\epsilon + b (x_1+1)-(x_1+1)(b - \beta)))$, and $m(x_1)'=\epsilon-b(x_1+1)^2-(x_1+1)(b-\beta))+(x_1+1)(-2bx_1-b-\beta)$. Since $m(0)=\epsilon-3b<0$, then $x_1=0$ is asymptotically stable. Accordingly, we can conclude the point $K_1$ is stable when $b<\beta+1$. 
When $b=\beta+1$, $K_1$ is unstable in accordance with the center manifold theorem whose derivation process is similar to the above analysis.

(2). For $K_2$: $(x,y,z,w)=(0,1,0,0)$, the corresponding eigenvalues of $J$ are
\begin{equation}
    \{\lambda_1,\lambda_2,\lambda_3\}=\{0,-1+b,-1+\epsilon\} .
\end{equation}
$K_2$ is unstable since $-1+b>0$.

(3). For $K_3$: $(x,y,z,w)=(0,0,1,0)$. Its corresponding eigenvalues of $J$ are
\begin{equation}
    \{\lambda_1,\lambda_2,\lambda_3\}=\{0,\epsilon,-\gamma\} .
\end{equation} 
When $\epsilon<0$, $K_3$ has an eigenvalue with zero real part and other eigenvalues with negative real part. According to the center manifold theorem, $K_3$ is stable. When $\epsilon>0$, $K_3$ is unstable because the eigenvalue $\epsilon$ has a positive real part. 

(4). For $K_4: (x,y,z,w)=(0,0,0,1)$. Its corresponding eigenvalues of $J$ are
\begin{equation}
    \{\lambda_1,\lambda_2,\lambda_3\}=\{-\epsilon,-\epsilon,-\epsilon\} .
\end{equation}
$K_4$ is stable when $\epsilon>0$ because all eigenvalues have negative real parts. $K_4$ is unstable when $\epsilon<0$ because all eigenvalues have positive real parts.

(5). For $K_5: (x,y,z,w)=(\epsilon,0,0,1-\epsilon)$. Its corresponding eigenvalues of $J$ are
\begin{equation}
    \{\lambda_1,\lambda_2,\lambda_3\}=\{0,\epsilon(-1+b-\beta),\epsilon(1-\epsilon)\} .
\end{equation}
When $0<\epsilon<1$ or $\epsilon<0$ and $b<1+\beta$, $K_5$ is unstable because one of its eigenvalues has a positive real part. When $\epsilon<0$ and $b\ge1+\beta$, $K_5$ has at least one eigenvalue with a zero real part and the others have negative real parts. According to the center manifold theorem, $K_5$ is unstable.

(6). For $K_6: (x,y,z,w)=(0,\epsilon,0,1-\epsilon)$. Its corresponding eigenvalues of $J$ are
\begin{equation}
    \{\lambda_1,\lambda_2,\lambda_3\}=\{0,\epsilon(-1+b),\epsilon(1-\epsilon)\} .
\end{equation}
When $\epsilon>0$, $K_6$ is unstable because eigenvalue $\epsilon(-1+b)>$. When $\epsilon<0$, there is one eigenvalue with a zero real part and two eigenvalues with negative real parts. According to the center manifold theorem, $K_6$ is unstable.

(7). For $K_7: (x,y,z,w)=(x^*,1-x^*,0,0)$. Its corresponding eigenvalues of $J$ are
\begin{equation}
    \{\lambda_1,\lambda_2,\lambda_3\}=\{0,-1+\epsilon,-1+b-\beta x^*\} .
\end{equation}
When $x^*>\frac{b-1}{\beta}$, namely $b<1+\beta$, there is one eigenvalue with a zero real part and others with negative real parts. According to the center manifold theorem, $K_7$ is stable. When $x^*<\frac{b-1}{\beta}$, $K_7$ is unstable because one of its eigenvalues has a positive real part.

(8). For $K_8: (x,y,z,w)=(x^*,\epsilon-x^*,0,1-\epsilon+x^*)$. Its corresponding eigenvalues of $J$ are
\begin{equation}
    \{\lambda_1,\lambda_2,\lambda_3\}=\{0,\epsilon-\epsilon ^2,-\epsilon+\beta-\beta x^*\} .
\end{equation}
When $\epsilon>0$, $K_8$ is unstable because $\epsilon-\epsilon ^2>0$. When $\epsilon<0$, $K_8$ is unstable because $-\epsilon+\beta-\beta x^*>0$. 

(9). For $K_9: (x,y,z,w)=(\frac{-1+b}{\beta},\frac{1-b+\beta}{\beta},0,0)$. Its corresponding eigenvalues of $J$ are
\begin{equation}
    \{\lambda_1,\lambda_2,\lambda_3\}=\{0,0,-1+\epsilon\} .
\end{equation}
$K_9$ exists only when $b<1+\beta$. When $K_9$ exists, there is one eigenvalue with a negative real part and two eigenvalues with zero real parts. According to the center manifold theorem, $k_9$ is unstable.

(10). For $K_{10}: (x,y,z,w) =
        (\frac{\epsilon}{b-\beta},0,\frac{\epsilon-\beta-\epsilon\beta}{(b-\beta)\gamma},1-\frac{\epsilon(\gamma+1+\beta-b)}{(b-\beta)\gamma})$. Its corresponding eigenvalues of $J$ are
\begin{equation}
\begin{split}
   & \{\lambda_1,\lambda_2,\lambda_3\} = \\
   & \{\!-\frac{\epsilon(-1+b-\beta)}{b-\beta}\!,\!
    -\frac{\epsilon(-1+b-\beta)}{b-\beta}\!,\!
    \epsilon \!+\! \frac{\epsilon^2(-1+b-\beta+\gamma)}{(b-\beta)\gamma}\!
    \}
\end{split} .
\end{equation}
$K_{10}$ exists when $1-\frac{\epsilon(\gamma+1+\beta-b)}{(b-\beta)\gamma})<1$, namely $b>\beta+\epsilon\frac{1-\gamma}{\gamma+\epsilon}$. Then its eigenvalue $\epsilon+\frac{\epsilon^2(-1+b-\beta+\gamma)}{(b-\beta)\gamma}>0$. Thus, $K_{10}$ is unstable.

(11). For $K_{11}: (x,y,z,w)=(\frac{(-1+b)\epsilon}{\beta},\frac{\epsilon-\beta+\beta\epsilon}{\beta},0,1-\epsilon)$. Its corresponding eigenvalues of $J$ are
\begin{equation}
    \{\lambda_1,\lambda_2,\lambda_3\}=\{0,0,
    \epsilon (1-\epsilon)
    \} .
\end{equation}
$K_{11}$ exists when $\epsilon>0$, then eigenvalue $\epsilon(1-\epsilon)>0$. Thus, $K_{11}$ is unstable.

(12). For $K_{12}: (x,y,z,w)=(\frac{\gamma}{1-b+\beta+\gamma},0,\frac{1-b+\beta}{1-b+\beta+\gamma},0)$. Its corresponding eigenvalues of $J$ are
\begin{equation}
\begin{split}
   & \{\lambda_1,\lambda_2,\lambda_3\}=\\
    &\{
    \frac{(1-b+\beta)\gamma}{1-b+\beta+\gamma}
    ,\frac{(1-b+\beta)\gamma}{1-b+\beta+\gamma},\epsilon+\frac{(-b+\beta)\gamma}{1-b+\beta+\gamma}
    \}
\end{split}.
\end{equation}
$K_{12}$ exists when $b<1+\beta$, then eigenvalue $\frac{(1-b+\beta)\gamma}{1-b+\beta+\gamma}>0$. Thus $K_{12}$ is unstable.   
\end{proof}
\clearpage
\section*{Supplementary Figures}
\begin{figure*}[!h]
    \centering
    \includegraphics[scale=0.66]{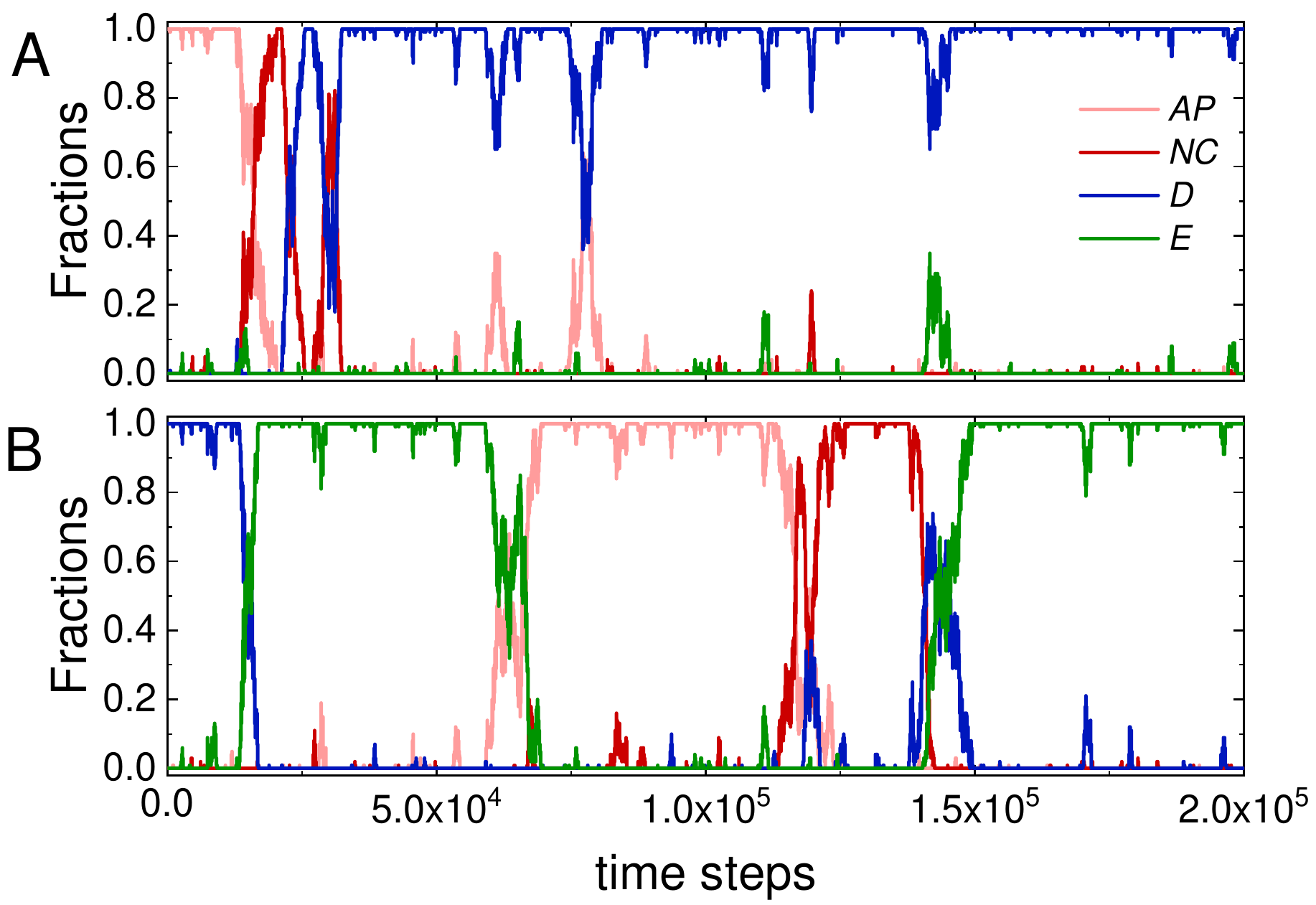}
    \caption{\textbf{Numerical simulation further demonstrates that the survival of altruistic punishment is due to the cyclic dominance between altruistic punishers, defectors, and exiters.} A. Defectors take over the whole population even if altruistic punishers initially dominate the population when the exiters' payoff is negative. B. Small but positive exiters' payoff enables the coexistence of altruistic punishers, non-punishing cooperators, defectors, and exiters through cyclic dominance. If defectors initially dominate the population, the mutated exiters invade the defectors, and after transient dynamics, the defectors finally give way to the exiters. When exiters dominate the population, altruistic punishment is less costly and cooperating is more valuable than exiting, and thus altruistic punishers take over the whole population. Thereafter, non-punishing cooperators dominate altruistic punishers and take over the whole population since altruistic punishers are less valuable than non-punishing cooperators. This proceeds until the dominance of non-punishing cooperators gives way to defectors again. The Parameter values are $b=1.5$, $\beta=0.3$, $\gamma=0.1$, $\mu=0.001$, $s=0.2$, $\epsilon=-0.2$ (A) and $\epsilon=0.2$ (B). }
    \label{fa1}
\end{figure*}

\begin{figure*}
    \centering
    \includegraphics[scale=0.6]{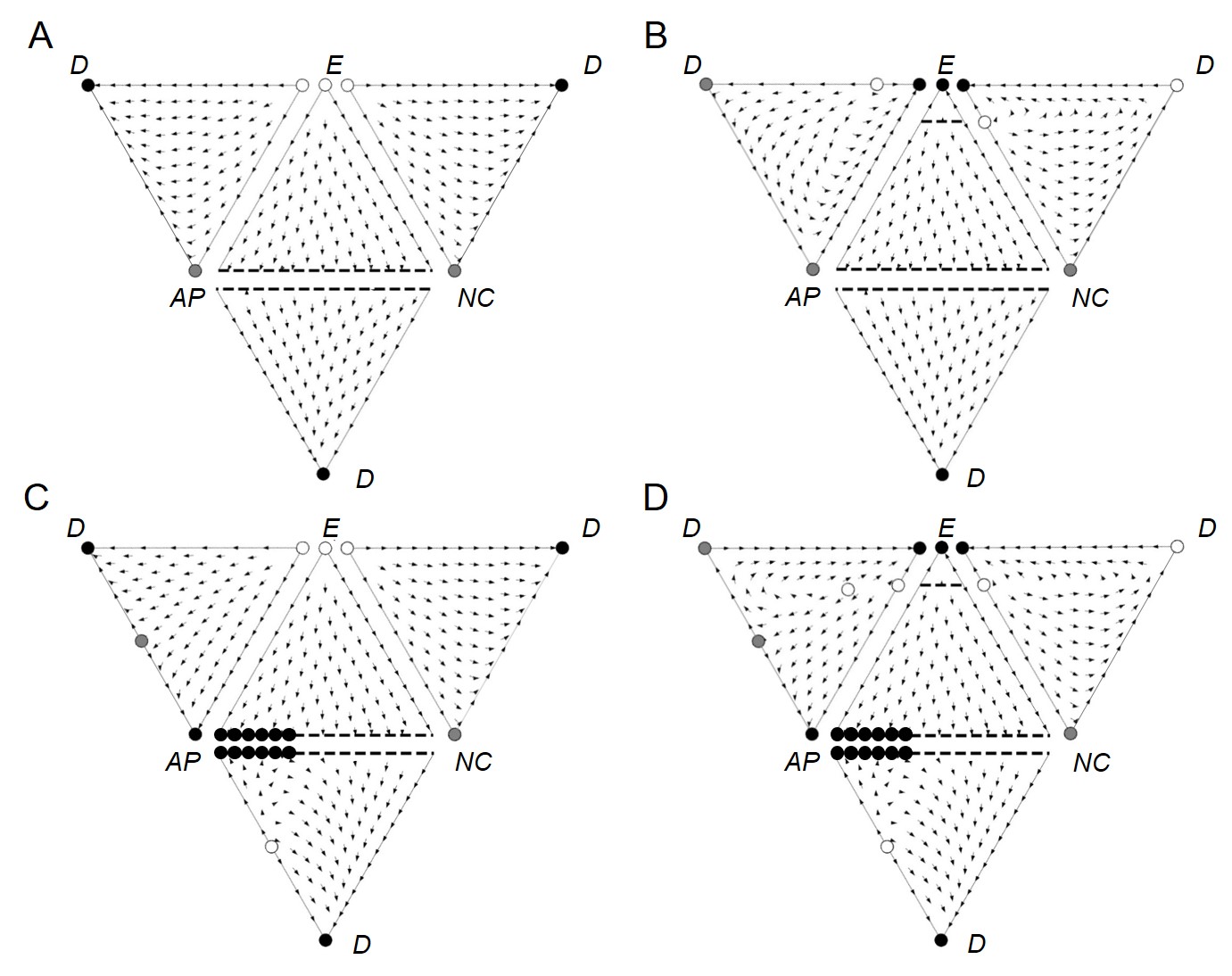}
    \caption{\textbf{Adding exit option destabilizes defection regardless of whether altruistic punishment can establish cooperation in an infinite population.} When the cost-to-effect ratio of altruistic punishment is insufficient to establish cooperation (top row), $b-\beta>1$, the monomorphic defecting equilibrium is replaced by the monomorphic exiting equilibrium for positive values of $\epsilon$. When the cost-to-effect ratio of altruistic punishment is capable of establishing cooperation (bottom row), $b-\beta<1$, the bi-stable equilibrium of the mixed altruistic punisher and non-punishing cooperator equilibrium and the monomorphic defecting equilibrium is replaced by the other bi-stable equilibrium between the mixed altruistic punisher and non-punishing cooperator equilibrium and the monomorphic exiting equilibrium for positive values of $\epsilon$. The dashed line on the $AP-NC$ edge indicates that all the points on this edge are unstable. The filled black circles, filled gray circles, and unfilled circles represent stable fixed points, saddle points, and unstable points, respectively. The parameters values are $\beta=0.3$, $\gamma=0.1$, $\epsilon=-0.2$ (left column), $\epsilon=0.2$ (right column), $b=1.5$ (top row), and $b=1.2$ (bottom row).}
    \label{fa2}
\end{figure*}

\end{document}